# Exponent equality for capture-zone scaling in island nucleation: Theory and application to organic films

Alberto Pimpinelli,[1,*] Levent Tumbek,[2] and Adolf Winkler[2,†]

[1]Rice Quantum Institute, Rice University, Houston, Texas 77005, USA

[2]Institute of Solid State Physics, Graz University of Technology, 8010 Graz, Austria

Abstract

It is known in thin-film deposition that the density of nucleated clusters $N$ varies with the deposition rate $R$ as a power law, $N \sim R^\alpha$. The exponent $\alpha$ is a function of the critical nucleus size $i$ in a way that changes with the aggregation limiting process active in a given system. We extend here to generic aggregation-limiting processes the derivation of the analytical capture-zone distribution function $P_\beta(s) = a_\beta \cdot s^\beta \cdot \exp(-b_\beta s^2)$ of Pimpinelli and Einstein [Phys. Rev. Lett. 99, 226102 (2007)]. We show that the parameter $\beta$ is generally related to the critical nucleus size $i$ and to the exponent $\alpha$ by the equality $\alpha(2\beta + d_f - 2) = 2i$ where $d_f$ is the fractal dimensionality of the clusters. This remarkable results allows one to measure $i$ with no a priori knowledge of the actual aggregation mechanism. We apply this equality to measuring the critical nucleus size in pentacene deposition on mica.

*Introduction.* When growing a thin film by depositing atoms or molecules on a substrate, one of the most easily accessible piece of information is the surface density of clusters—also called islands—after a given deposition time. As Venables and coworkers [1] showed many years ago the island density $N$ is a function of the substrate temperature $T$ and of the deposition rate $R$, and it depends on a material parameter, the size of the critical nucleus, $i$. The latter is the number of atoms that are part of the largest unstable cluster: a cluster of size $i+1$ can only grow—through capture of diffusing monomers—while smaller cluster can both grow and dissolve. The critical nucleus size itself is in reality a function of $T$ and $R$, but simulations and actual experiments show that a real—or realistic—system behaves in some range of values of the external parameters *as though* a given value of $i$ was selected. In particular, Venables and coworkers showed that the island density at fixed substrate temperature behaves as a power-law function of the deposition rate, $N \sim R^\alpha$, where the exponent $\alpha$ depends on $i$. Measuring the island density as a function of the deposition rate allows thus one to obtain the value of $i$ for given experimental conditions.

Measuring $N$ was made possible by observation techniques such as Scanning Tunneling Microscopy (STM) or Atomic Force Microscopy (AFM). The same techniques allow one to count the number of monomers in each cluster and thus obtain the island size distribution (ISD) for given $T$ and $R$. Simulations have shown [2] that the ISD can be quantitatively described by an ad hoc analytic expression which contains a single parameter, the critical cluster size $i$. Therefore, that expression has been widely used to extract $i$ from STM or AFM images.

More recently, Pimpinelli and Einstein (PE) [3] have proposed an alternative analytic approach for extracting values of $i$ based on the capture zone distribution (CZD). A capture zone, as the name suggests, is the region of the substrate around an island that collects monomers that are most likely to be captured by that island. Capture zones are approximated (see the inset in Fig. 2 below) by the more manageable Voronoi polygons—regions of the plane whose points are closest to their center than to any other point. The CZD is then the statistical distribution of the sizes (areas) of the capture zones. The idea of using CZD in connection with island nucleation dates back to the 90s [4]. The novel feature of PE's proposal was to provide a prescription for computing the analytical—though approximate—form of the CZD depending on a single parameter, which turned out to be precisely $i$, the size of the critical nucleus. The analytical form coincides with the so-called Generalized Wigner distribution (GWD) previously used for describing, among other things, fluctuations of crystal steps [5]. The agreement between the GWD and CZD has been checked against results from kinetic Monte Carlo (kMC) simulations by several authors. A thorough discussion of kMC simulations can be found in Ref. [6], which concludes that the GWD is indeed an excellent quantitative approximation to the CZD.

The original PE derivation was made for nucleation—or aggregation—limited by surface diffusion, that we will call DLA. This is also the only aggregation mechanism for which an analytical approximation to the ISD exists [2]. We will show here that PE's approach can be generalized to any kind of aggregation mechanism, which makes it a much more powerful tool than the ISD. In fact, we will show that coupling the measurements of the island density as a function of the deposition rate with measurement of the CZD allows one to extract the critical nucleus size $i$ without the need of any assumptions concerning the nucleation mechanism.

We will apply these novel results to deposition, diffusion and aggregation of pentacene molecules on sputter amorphized mica surfaces. This system exhibits a crossover between different nucleation regimes. We will show how PE's approach allows one to determine $i$ without *a priori* assumptions on the aggregation process.

*Theory*. One of the most interesting aspects of PE´s approach is the connection between the form of the CZD and nucleation: the small-area behavior of the distribution is directly dictated by creation of new CZs when new islands are nucleated. PE conjectured a proportionality relation between the probability $P(s)$ of finding a given value of the dimensionless area $s = A/<A>$ of a CZ (at small $s$), and the probability of nucleating a new island. The latter is in turn proportional to $\int dr\, r\, [n(r)]^{i+1}$, $n$ being the density of diffusing monomers, and the integral being computed over half the average distance between islands, $L$ [7]. The precise functional form of the CZD depends therefore on the monomer density inside the CZ or, more precisely, on the way the monomer density scales with the area of the CZ. This is in turn determined by the characteristics of the diffusion and aggregation process involved. For instance, if aggregation is limited by diffusion (DLA) as in Ref. [3], so that attachment of monomers at the rim of the islands is fast compared to diffusion, the monomer density $n$ must vanish at the island edge. The average monomer density can be then shown to scale as the area $A = L^2$ of the capture zone, within logarithmic corrections [7]. The nucleation probability scales thus as $\int dr\, r\, [n(r)]^{i+1} \sim L^{2(i+2)} = A^{(i+2)}$. The corresponding CZD has therefore the form of a GWD, $P_\beta(s) = a_\beta \cdot s^\beta \cdot \exp(-b_\beta s^2)$, where the parameter $\beta$ reads $\beta_{DLA} = i+2$.

However, different types of aggregation regimes are observed: besides diffusion, nucleation can be limited by attachment barriers, reactions, blocking impurities, desorption, just to name a few. Anisotropies in diffusion and/or attachment can also lead to different nucleation regimes. Such differences manifest themselves in the power-law relation $N \sim R^\alpha$: the way the exponent $\alpha$ depends on $i$ changes with the various limiting processes. This happens because, in the steady state where most islands have formed, the monomer density is fixed by the balance between deposition and capture by islands—at least, when monomer desorption is negligible. Assuming that just a single aggregation mechanism is at work—a quite strong assumption, in fact—the monomer density $n$ will scale as a power of the island density $N$, $n \sim R/D\, N^{-\gamma}$ [8], where $\gamma$ is characteristic of the limiting process: $\gamma = 1$ for isotropic diffusion, $\gamma = 2$ for strongly anisotropic (1D) diffusion, $\gamma = 1/a$ for restricted monomer diffusion, where $a$ is an *a priori* unknown function of the concentration of blocking impurities [8]. In the case of attachment-limited aggregation (ALA), as shown by Kandel [9] the monomer density scales as the distance between islands, so that $\gamma = \frac{1}{2}$. The scaling of the island density with deposition rate in the saturation regime—where the island density has its maximum—is then found balancing the nucleation rate $D/R\, n\, n^i$ against island coalescence (see Ref. [8] for details)

$$N/\theta_c \approx D/R\, nn^i \qquad (1)$$

where $\theta_c$ is the surface coverage when islands come into contact, $D$ the monomer diffusion coefficient, and $n^i$ is a mean-field approximation to the density of critical nuclei of size $i$ (Walton relation [10]). A temperature-dependent coefficient has been omitted from Eq. (1). As shown in Ref. [8], $\theta_c$ is in special situations (e.g. when clusters are fractal) a function of the island density itself. Assuming that $\theta_c \sim N^\delta$ and replacing $n \sim R/D\, N^{-\gamma}$ into Eq. (1) yields $N \sim R^\alpha$, with $\alpha = i/(\gamma i+1+\gamma-\delta)$. In the case of DLA and isotropic diffusion ($\gamma = 1$) this leads to the well-known relationship $\alpha = i/(i+2)$, with $\delta = 0$. For attachment-limited aggregation (($\gamma = ½$ and $\delta = 0$) one finds $\alpha = 2i/(i+3)$ [9,11].

What about the CZD? Consider PE's prescription for the small-$s$ behavior of $P(s)$: it must follow from equating $P(s)$ and $\int dr\, r\, [n(r)]^{i+1}$. Substituting $n \sim R/D\, N^{-\gamma}$ and $N \sim L^{-2} \sim A^{-1}$ yields $P(s) \sim s^{[1+\gamma(i+1)]}$ at small $s$. Factoring in the Gaussian decay [3] at large $s$ yields a GWD $P_\beta(s) = a_\beta \cdot s^\beta \cdot \exp(-b_\beta s^2)$ where

$$\beta = \gamma i+1+\gamma. \qquad (2)$$

As a consequence, the equality holds

$$\alpha\,(\beta - \delta) = i. \qquad (3)$$

Note that, when $\delta = 0$ Eq. (3) reduces to

$$\alpha\,\beta = i. \qquad (4)$$

Eq. (3) also holds as an approximate identity valid when $\delta \ll \beta$, which is true in most situations of practical interest. As an example, consider a (possibly) fractal island growing in diffusion-limited conditions (DLA). The "mass" $M$ (number of monomers) of an island at time $t$ obeys the equation $M = j_{DLA}\, t$, where $j_{DLA} = Dn$ is the monomer current to the island edge. For 2D growth, the island mass $M$ and its radius $r$ are related by $M = (r/a)^{d_f}$, where $d_f$ is the—possibly fractal—dimensionality of the island, and $a$ a molecular length scale. The surface coverage at which two islands come into contact, $\theta_c = R\, t_c$ is found by letting $r = 1/N^{1/2}$ at $t = t_c$. Therefore, $Dn \sim R\, N^{-1}$ ($\gamma = 1$ for DLA in 2D), so that $R\, t_c \sim N^{1-d_f/2}$ and $\delta = 1 - d_f/2$. Hence, for compact islands ($d_f = 2$), $\delta = 0$, while $\delta = 0.15$ for fractal islands with $d_f = 1.7$.

The same result is found for islands growing in attachment-limited conditions (attachment-limited aggregation, or ALA). In this case, $M = j_{ALA}\, t$, where $j_{ALA} = 2\pi r\, kn$, and $k$ is an attachment kinetic coefficient. As shown by Kandel [9], the monomer density scales as the distance between islands, $kn \sim R\, N^{-1/2}$ ($\gamma = \tfrac{1}{2}$) so that $R\, t_c \sim N^{1-d_f/2}$ and again $\delta = 1 - d_f/2$. We can therefore generalize Eq. (3) to fractal islands:

$$\alpha\,(2\beta + d_f - 2) = 2i. \qquad (5)$$

As a byproduct of Eq. (3) the exponent $\beta_{ALA}$ of the corresponding GWD can be computed for compact islands, yielding $\beta_{ALA} = (i+3)/2$. The latter result is new.

Equations (2) to (5) are the main theoretical results of the present Letter. They are indeed remarkable, because they imply that measuring $\alpha$ and $\beta$ (and possibly the fractal dimensionality of the clusters) allows one to measure the critical nucleus size, *without any knowledge of the aggregation mechanism*.

*Experiments.* In order to show how this theoretical approach can be useful for interpreting experiments, we have investigated island nucleation in a typical organic system, pentacene deposited on amorphized mica. Pentacene (5A) was evaporated on sputter amorphized mica at room temperature under UHV conditions ( $1\times10^{-8}$ mbar) from a Knudsen cell. Under these conditions islands are formed which are composed of standing molecules. The film morphology was determined by ex-situ atomic force microscopy (AFM). The growth behavior is similar to that for the system p-hexaphenyl (6P) on sputtered mica [12,13,14]. That system was shown not to conform to standard reversible DLA [15]. As stated above, (reversible) DLA is characterized by an island density $N$ that, in the sub-monolayer regime, depends on the deposition rate $R$ as $N \sim R^\alpha$, with $\alpha = i/(i+2)$. It is readily seen that $\alpha$ can only take values between 0.33 and 1 in DLA However, deposition experiments of 6P molecules on mica resulted in exponents $\alpha$ larger than one [15], inconsistent with DLA. The growth kinetics of pentacene (5A), cannot be described simply by reversible DLA either, shown in Fig. 1. In particular, a crossover is observed in the ln$N$ vs. ln$R$ plot between a value of the exponent $\alpha = 0.8 \pm 0.1$ at low, and $\alpha = 1.3 \pm 0.1$ at high deposition rate. We have then measured the CZD in both the low (Fig. 2a) and high $R$ range (Fig. 2b), and determined $\beta = 5.0 \pm 0.5$ for the low and $\beta = 4.0 \pm 0.5$ for the high $R$ range, respectively. This allows us to estimate the critical nucleus size using Eq. (4), assuming that $\delta$ is either vanishing or small. We find $i = 4.0 \pm 0.9$ in the low $R$ regime, and $i = 5.2 \pm 0.9$ in the high $R$ one.

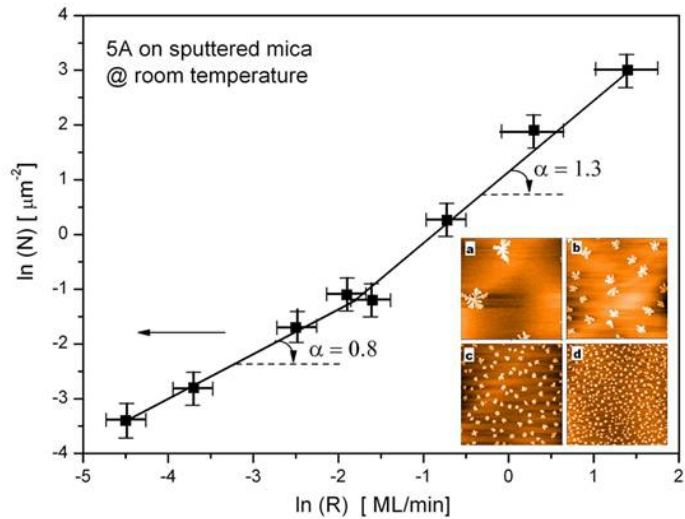

FIG.1. Island density $N$ as a function of deposition rate $R$ at 300 K. The slope at low rate is $\alpha = 0.8 \pm 0.1$, at high rate $\alpha = 1.3 \pm 0.1$. The inset shows exemplary AFM images (8 μm x 8 μm) for different deposition rates. a: 0.01 ML/min, b: 0.15 ML/min, c: 0.48 ML/min, d: 1.37 ML/min.

The traditional approach would be to use the measured value of $\alpha$ to *surmise* the aggregation regime, and then compute $i$ from the theoretical expressions for $\alpha$ and $\beta$. Thus, the value measured at high deposition rate is inconsistent with DLA, and possibly consistent with ALA. Using now $\alpha = 2i/(i+3)$ and the experimental value $\alpha = 1.3 \pm 0.1$ yield $i = 5.6 \pm 1.4$. Using $\beta = (i+3)/2$, ($\beta = 4.0 \pm 0.5$) yields $i = 5.0 \pm 1.0$. The value from Eq. (4) is right in between, and consistent with both values within error bars.

At low deposition rate, the measured $\alpha = 0.8 \pm 0.1$ is (possibly) consistent with both DLA and ALA. Assuming that DLA holds, $\alpha = i/(i+2)$ yields an estimate for $i$ that varies between 4.7 and 18, while $\beta = i+2$ gives $i = 3 \pm 0.5$. Assuming ALA, $\alpha = 2i/(i+3)$ yields $i = 2.0 \pm 0.4$, while $\beta = 5.0 \pm 0.5$ and $\beta = (i+3)/2$ gives $i = 7 \pm 1.0$. The value $i = 4.0 \pm 0.9$ as obtained by Eq. (4) is roughly consistent with nucleation being in a DLA regime at small deposition rates. We stress, however, that the values obtained from Eq. (4) do not assume any given aggregation mechanism, and are unbiased measures of the size of the critical nucleus size.

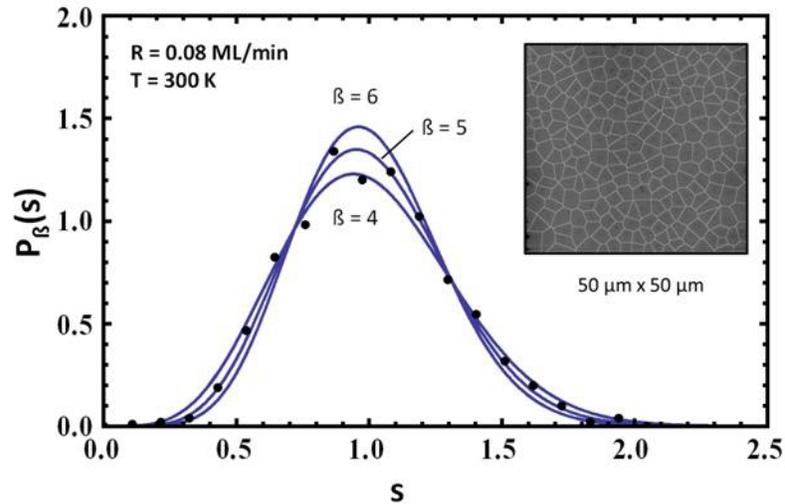

FIG. 2a. Capture zone distribution (black dots) obtained by summing over 5 different AFM images of a 5A film deposited on sputtered mica with a low deposition rate $R = 0.08$ ML/min at room

temperature. The curves were calculated using the fit function $P_\beta(s)$ from PE [3]. The best fit yields $\beta = 5.0 \pm 0.5$. The inset shows a representative Voronoi tessellation (50 μm x 50 μm).

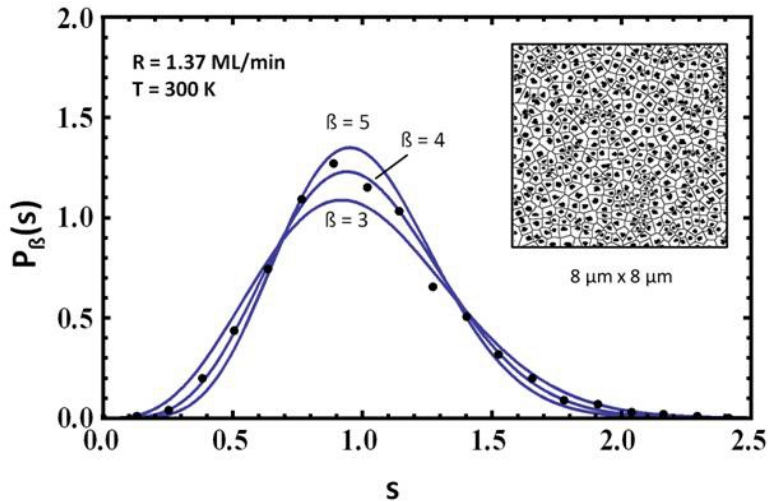

FIG. 2b. Capture zone distribution (black dots) obtained by summing over 5 different AFM images of a 5A film deposited on sputtered mica with a high deposition rate $R = 1.37$ ML/min at room temperature. The curves are calculated using the fit function $P_\beta(s)$ from PE [3]. The best fit yields $\beta = 4.0 \pm 0.5$. The inset shows a representative Voronoi tessellation (8 μm x 8μm).

*Conclusions.* The PE proposal that the GWD be used to describe the CZD in the sub-monolayer regime shows here all its strength: even though it is far from being an exact result, it has the merit of deriving from a formal mathematical argument that allows it to be extended to any aggregation-limiting process. This is not the case for the ISD, whose analytic form was derived from fits to Monte Carlo simulations of reversible DLA. Nothing is known to what the ISD should be in other instances, e.g. for ALA. The present extension of the GWD to any nucleation process has of course its own limitations. For instance, it requires a single nucleation process to be active, and it is not able to provide, if not approximately, the correct value of the parameter $\beta$ in the cross-over region if two nucleation processes are simultaneously present. More experimental and computational work is needed to investigate the range of applicability of the GWD, and the validity of the exponent equalities that we established here.

LT and AW were supported by the Austrian Science Fund, FWF, Proj. No. P23530.

\* [ap19@rice.edu](mailto:ap19@rice.edu)

† [a.winkler@tugraz.at](mailto:a.winkler@tugraz.at)